# Facing Complexity:
# Prediction vs. Adaptation

Carlos Gershenson
Departamento de Ciencias de la Computación
Instituto de Investigaciones en Matemáticas Aplicadas y en Sistemas
Universidad Nacional Autónoma de México
cgg@unam.mx

*Abstract*: One of the presuppositions of science since the times of Galileo, Newton, Laplace, and Descartes has been the predictability of the world. This idea has strongly influenced scientific and technological models. However, in recent decades, chaos and complexity have shown that not every phenomenon is predictable, even if it is deterministic. If a problem space is predictable, in theory we can find a solution via optimization. Nevertheless, if a problem space is not predictable, or it changes too fast, very probably optimization will offer obsolete solutions. This occurs often when the immediate solution affects the problem itself. An alternative is found in adaptation. An adaptive system will be able to find by itself new solutions for unforeseen situations.

## Introduction

The scientific method, since the epoch of Galileo, Newton, Laplace and Descartes, has contributed considerably to the increase of human knowledge and to technological development. Still, its success does not imply that it cannot or should not be criticized, especially considering the differences between the context in which it was proposed and the present context.

The traditional scientific method isolates and simplifies a problem to proceed to mathematical formalization as a tool to finding a solution. This method is useful for several problems, such as building traditional bridges, automobiles, and computers, since there are relatively few variables to be considered and these do not affect each other too much, i.e. they can be treated in isolation. Moreover, the specifications of these problems do not change: gravity, combustion, electrical properties, etc. are constant.

There are two ways in which the traditional method becomes inadequate. On the one hand, if there are many variables that determine a system, their combinatory explosion prevents the finding of a solution via analysis or exhaustive search. Optimization techniques can be useful to find adequate solutions. On the other hand, if the problem itself changes, the solution will probably be obsolete. If the problem changes faster than it can be optimized, a different approach is required. This is because it is not possible to predict the future of the problem and traditional techniques become inadequate. The traditional method requires the complete prespecification of a problem to find solutions (from initial and boundary conditions) (Gershenson, 2011a). However, we often encounter problems that cannot be prespecified, especially when solutions themselves change the problem.

# *The Limits of Prediction*

Since at least the late nineteenth century, the scientific presupposition concerning the predictability of the world was questioned (Morin, 2007), starting with the three-body problem studied by Poincaré. Nevertheless, today many a scientist still assumes that the world is and must be predictable.

The predictability presupposition was proven mistaken with the study of deterministic chaos. It is logical to think that a deterministic system is predictable, which led Laplace to propose his famous "daemon". Laplace hypothesized that if a superior intelligence could have access to all of the positions and momentums of all particles in the universe, Newton's laws (which are deterministic and reversible) could be used to know all past and future events in the universe. There are several problems with the worldview exemplified by Laplace's daemon:

1. Even with a complete description of all elementary particles of the universe (whatever these may be), phenomena at different scales cannot be predicted from this description of the universe. Life, mind, dreams, imagination, Rachmaninov's $2^{nd}$ piano concerto, money, a Murakami novel, a revolution. All of these phenomena are *real* and have a causal effect on the physical world, but cannot be described in terms of the laws of elementary particles. The universe is not *reducible*.

2. A complete model of the universe must contain the model itself. This leads to a paradox, similar to Russell's. If the model contains the universe, but is a part of the universe, it has to contain itself infinitely.

3. Irreversibility in thermodynamics showed that it is not possible to deduce all past events. For example, if there are two states that lead to the same third state, once being in the third state it is not possible to determine which of the two states led to it.

4. Determinism does not imply predictability. Predictability is limited in *chaotic* systems, which are "sensitive to initial conditions". In chaotic systems, very similar initial states can lead to very different future states. For example, a variable with an initial value of 3.3333333333 can lead to a final value of 0.25, while an initial value of 3.3333333334 can lead to a final value of -1.6793. Independently of how much precision is considered, very small differences will lead to very large differences, since trajectories diverge exponentially (this can be formally measured with Lyapunov exponents). This sensitivity to initial conditions is a characteristic of *deterministic chaos* (Elert, 1995; Gershenson, 2002a). Lacking an infinite precision, even with complete information about a state and functioning of a system, its predictability can be limited.

A classical example of chaos is found in weather forecast. This is limited not because atmospheric dynamics are unknown to meteorologists, but because of the inherent chaos present in the atmospheric dynamics. Even if precision is increased, predictions cannot be made with a high confidence more than a few days in advance.

Another example of a system with a limited predictability is road traffic. Vehicle dynamics can be described with classical mechanics. However, there are many additional factors that affect vehicle movement, such as driving conditions (wet floor, poor visibility, road works), the state of the driver (distracted, tense, in a hurry, texting, sleepy, under the effects of certain substances), pedestrians (children playing, jaywalkers, street merchants), etc. It is possible to attempt to predict the future position of a vehicle, but it will be limited to not much more than two minutes. Minimal changes in the predicted trajectory of a vehicle can lead to major effects in the traffic of a whole city. This is due to the large amount of interactions between each vehicle and its environment: other vehicles, pedestrians, traffic lights, etc. (Gershenson, 2005; 2007). These relevant interactions are one of the main characteristics of *complex systems* (Bar-Yam, 1997; Gershenson & Heylighen, 2005).

## *Complexity*

It is difficult to define complexity precisely, since it can be found everywhere. Etymologically, complexity comes from the Latin *plexus*, which means interwoven. In other words, a complex system is difficult to separate. This separation is difficult because the *interactions* between the components of the system are relevant, as the future of each element depends on the state of other elements. Since interactions generate novel information, which is not present in initial nor in boundary conditions, the future of complex systems cannot be *reduced* to the isolated dynamics of its components. Traditional science attempts precisely this, to simplify and isolate in order to predict, reducing the behavior of a system to that of its components. But a (reductionist) model that does not take into account relevant interactions will not be useful, as predictions will probably be mistaken. Moreover, the behavior of the system will be difficult to understand if this is reduced to the behavior of the parts, precisely because relevant interactions are not considered.

Examples of complex systems are everywhere: cells, brains, cities, Internet, stock markets, insect colonies, ecosystems, biospheres. All of these systems consist of elements that interact to produce a system behavior that depends on the elements and their interactions. For example, cells are made by molecules. Cells are living, while molecules are not. Where do life and its organization come from? These come precisely from the *interactions* between molecules. A similar example: brains are composed by neurons and molecules. Brains are capable of reasoning, imagination, consciousness, etc. These properties are not present in the components. Where do they come from? From the interactions. It is because of the relevance of interactions that the behavior of a system cannot be reduced to the behavior of its parts. Interactions generate novel information that is not present in the parts but is essential for their behavior, and thus for the system.

A classic example of complexity can be seen with Conway's "Game of Life" (Berlekamp et al., 1982). This consists of a lattice, in which each "cell" can take one of two values: 1 ("alive") or 0 ("dead"). The state of each cell depends on its eight closest neighbors: if there are less than two neighbors around a living cell, this dies out of "loneliness". If there are more than three, it also dies, because of "overpopulation". If there are two or three living neighbors, the cell will remain alive. If around a dead cell there are precisely three living cells, a new cell is "born". These simple rules produce an impressive complexity (see Figure 1). There are certain stable structures that can emerge from random initial conditions. There are also oscillatory structures that repeat a dynamical pattern with a specific period. There are also moving structures, which travel across the lattice, until they encounter other structures, with which they interact. There are oscillatory structures that produce moving structures. There is a richness of dynamical patterns that has yet to be exhausted. Moreover, it is possible to implement a universal Turing machine (capable of computing any computable function) in the Game of Life. Could all of this richness be predicted from the rules of the game? Simple rules produce complex behaviors and patterns through local interactions. The large scale properties cannot be known a priori, since these are *computationally irreducible* (Wolfram, 2002).

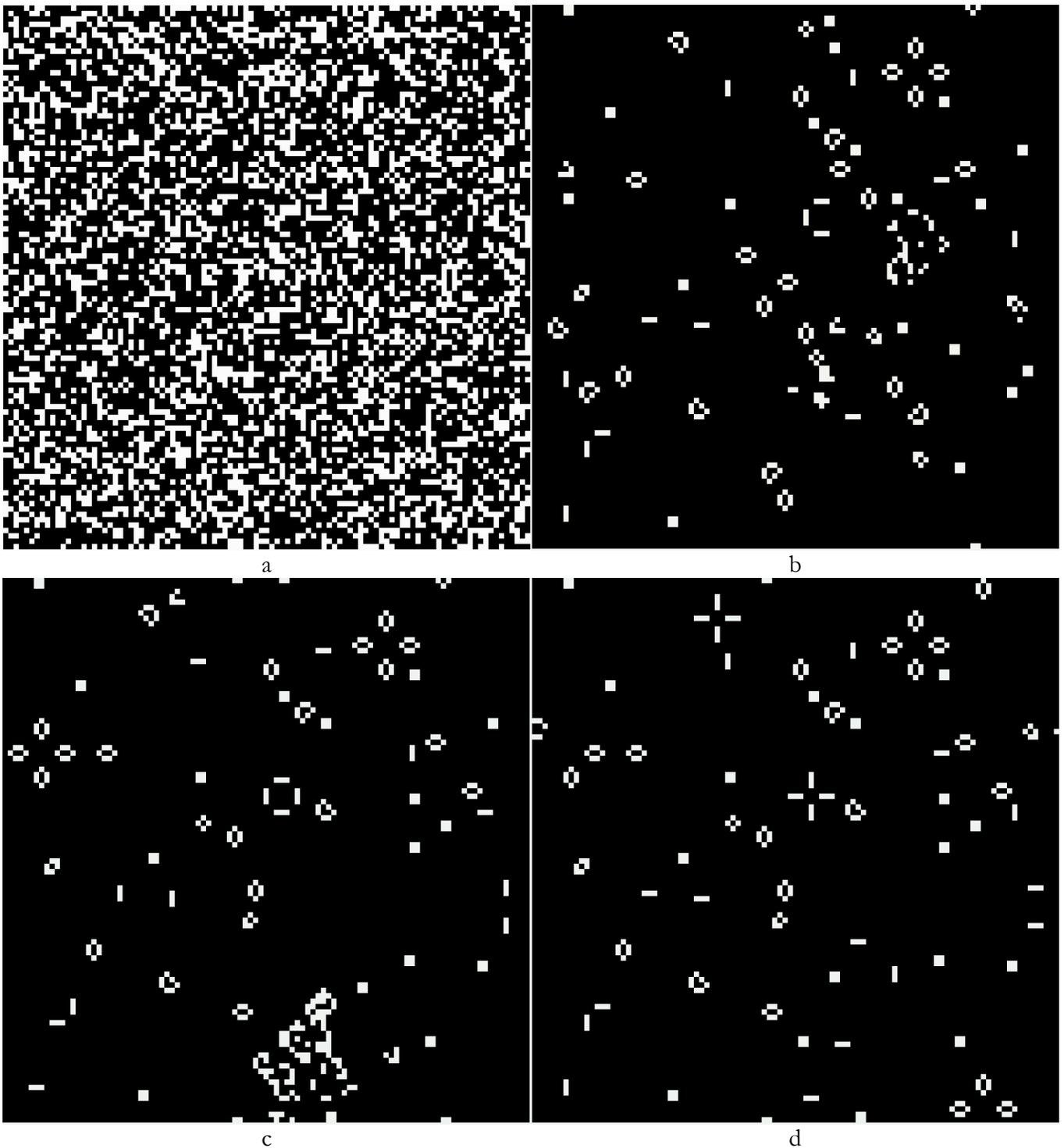

Figure 1. Evolution of the Game of Life from a random initial condition (a), where white cells are "alive" and black cells are "dead". After 410 steps (b), certain stable structures have been formed, but there are still some active zones. After 861 steps (c), some structures have been destroyed and some new ones have been created. Activity continues in the lower part of the lattice. After 1416 steps (d), the dynamics is periodic, with stable and oscillatory structures. Images created with NetLogo (Wilensky, 1999).

Another example of the relevance of interactions can be seen with elementary cellular automata (Wolfram, 1986; 2002; Wuensche & Lesser, 1992). The Game of Life is a two-dimensional cellular automaton, since cells are arranged on a plane. Elementary cellular automata are unidimensional. They consist of an array of cells, each of which can take one of two values: zero or one. The state of each cell depends on its previous state and on the previous state of its two closest neighbors. This is determined by a lookup table, which consists of the eight possible combinations of zeroes and ones on three cells (111, 110, 101, 100, 011, 010, 001, 000), and an assigned value (zero or one) for each combination. Having eight combinations and two possible values, there are $2^8=256$ different "rules" (11111111, 11111110, 11111101, ..., 00000000). Transforming these strings to base ten, these can be referred with a number, e.g. rule 10101010 corresponds to $2^7+2^5+2^3+2^1=128+32+8+2=170$. While there are 256 rules, many of them are equivalent, so there are only 88 "clusters" with different dynamics. There are rules that produce simple, repetitive patterns (e.g. rules 254, 250). Other rules produce nested structures (e.g. rules 90, 22). There are also rules that produce pseudorandom patterns (e.g. rules 30, 45). Finally, there are also rules that produce localized structures (e.g. rule 110). These cases are illustrated in Figure 2. Rule 110 is an interesting case. Similarly to the Game of Life, there are structures that persist in time and glide in space. When structures collide, they interact and may be transformed. It should be noted that there are interactions at different scales: between cells and between structures. It has been shown that rule 110 can also implement a universal Turing machine. Being such a simple system composed only by zeroes and ones and determined by eight bits, it has an immense potential. Where does this complexity come from? From *interactions*.

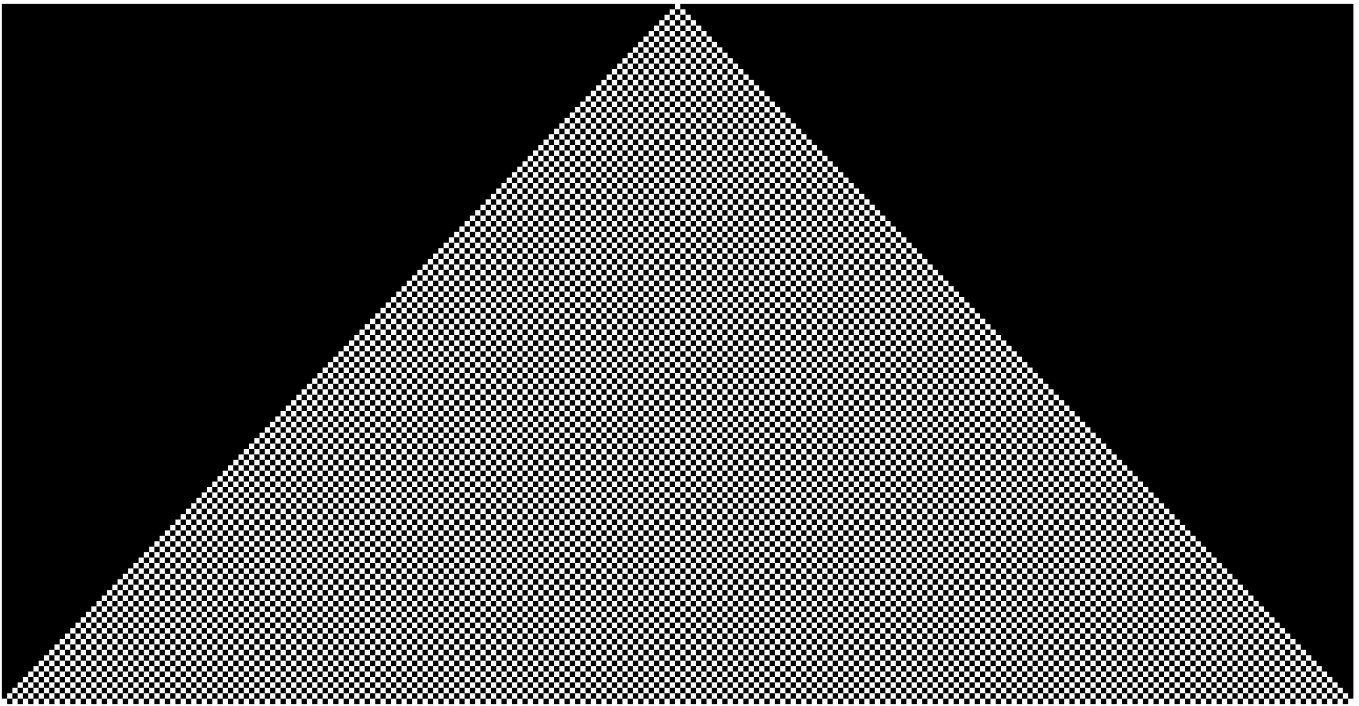

a

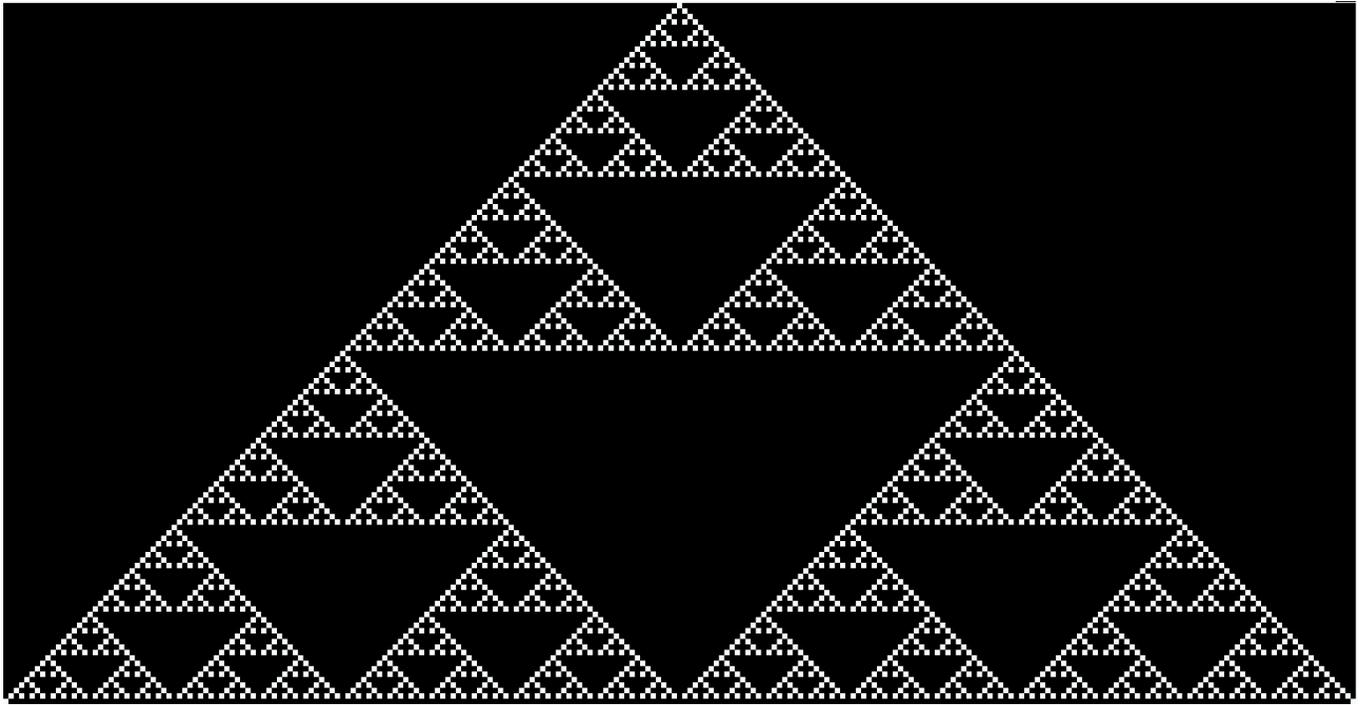

b

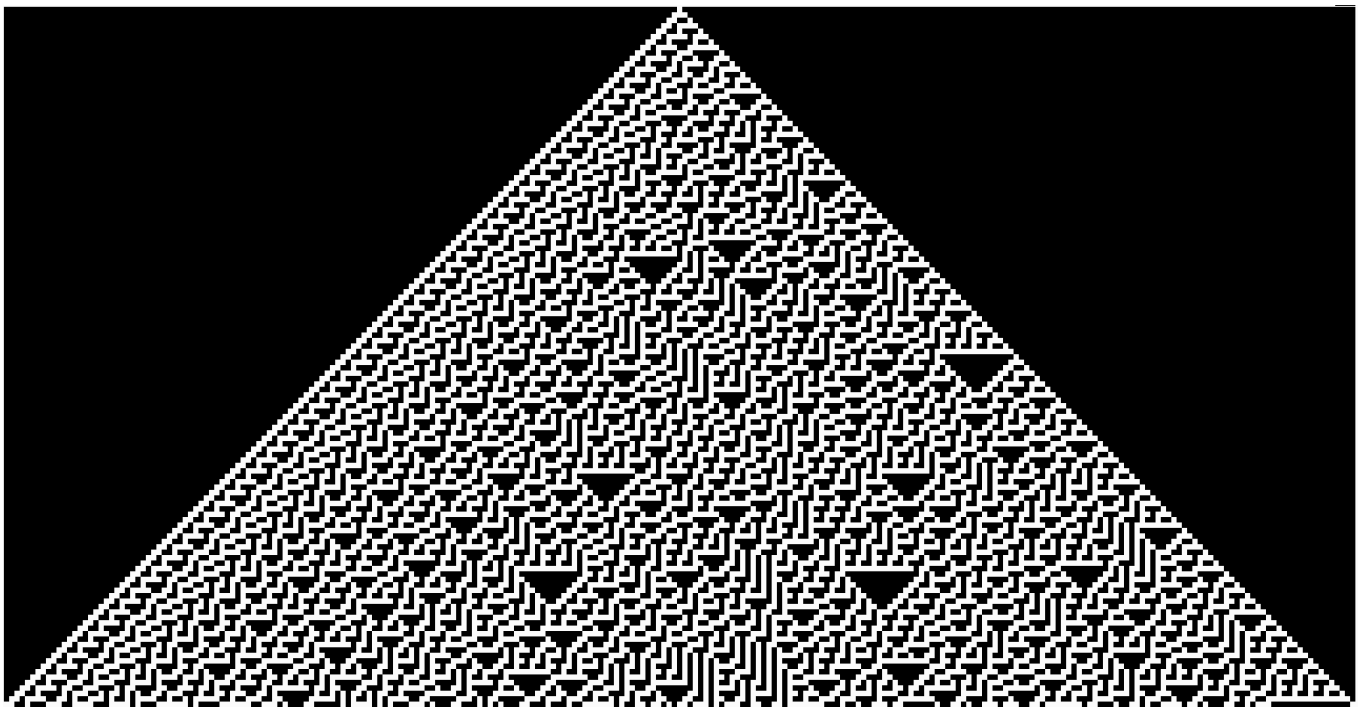

c

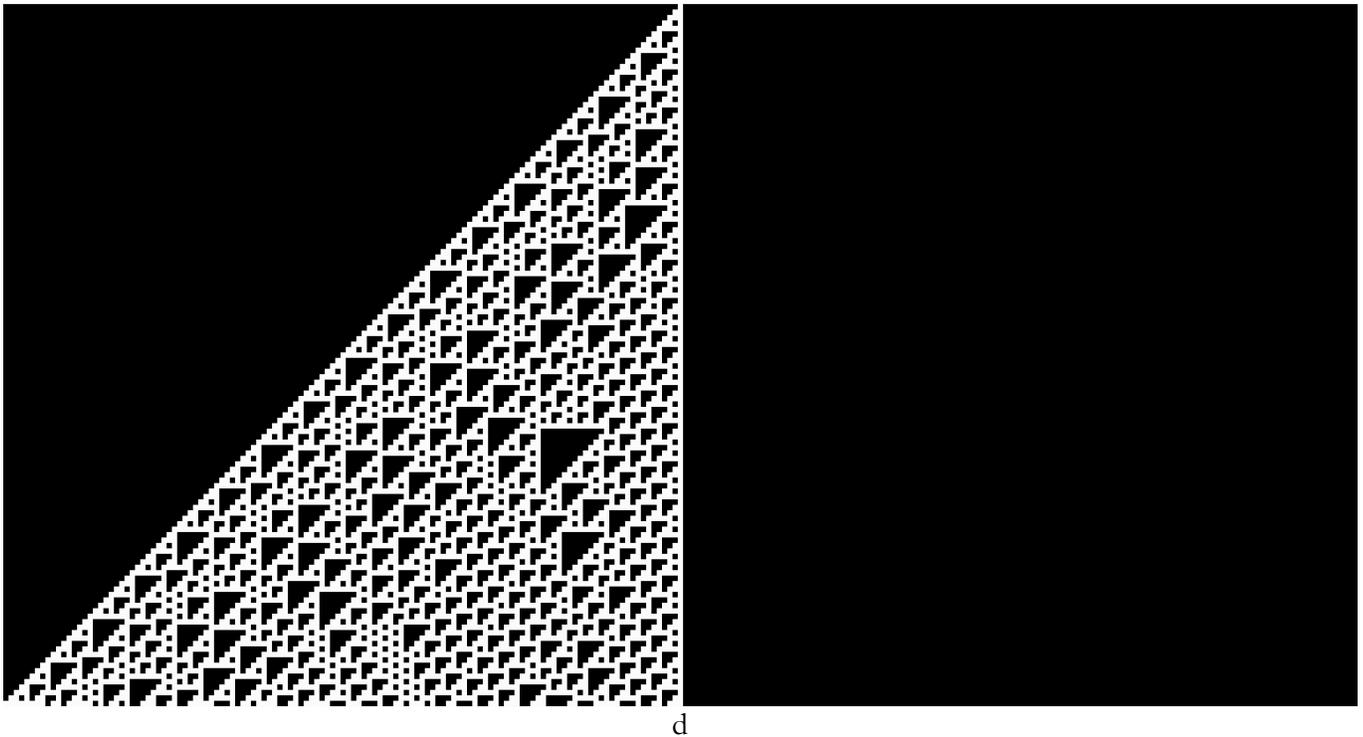
Figure 2. Examples of different classes of elementary cellular automata, with an initial state of a single active cell on the top row, time flows downwards. Rule 250 (a) produces regular patterns (class I). Rule 90 (b) produces nested patterns. Rule 30 (c) produces pseudorandomness (class III). Rule 110 (d) produces localized structures (class IV). Images created with NetLogo (Wilensky, 1999).

Complexity carries with it a lack of predictability different to that of chaotic systems, i.e. sensitivity to initial conditions. In the case of complexity, the lack of predictability is due to relevant interactions and novel information created by them. This novel information can also consist of new variables. Since precise interactions and novel variables cannot be prespecified, solutions to complex problems cannot be found *a priori*. For example, the future state of a system as "simple" as rule 110 cannot be known without actually running it. The emergent structures of the Game of Life cannot be predicted from its simple rules if the system has not been run. Also, when a problem changes the solution that had been found for it becomes obsolete. Then the problem space can be called *non-stationary* (Gershenson, 2007). Almost all problems fall within this category, since all systems in the real world are *open*. In other words, systems interact with their environment. They are not completely isolated. Thus, systems become unpredictable on the long run. This is because an open environment lacks predictability, and an open system is affected eventually by the unpredictable changes of its environment.

What can be done to solve problems with a non-stationary space? A good starting point is to take inspiration in nature, since living systems are constantly solving a non-stationary problem: survival.

## *Adaptation*

*Adaptation* (Holland, 1995) is the ability of a system to change its behavior when facing a perturbation. Living systems have to adapt constantly to changes in their environment, so they are a source of inspiration for building biomimetic adaptive systems.

The difference between adaptation and prediction is that the latter tries to act before a perturbation affects the expected behavior of a system. Certainly, it is desirable to predict perturbations, since these can affect negatively or even destroy a system. However, as it has been shown, it is not possible to predict all future interactions of a system. This is why it becomes necessary to build systems that are able to adapt, since there will be unexpected situations. An adaptive system will be able to respond to the unexpected, to a certain degree, without the need of human intervention.

It can be said that adaptation is a type of creativity (Kauffman, 2008). Adaptive systems can create novel solutions. This is necessary if systems are expected to face a complex and unpredictable environment.

There are several techniques to build adaptive systems. One of them is to use the concept of self-organization (Gershenson, 2007).

## *Self-organization*

A system can be described as self-organizing if its elements interact so that the behavior of the system is a product mainly of these interactions, not from a single element or from an external source. All of the examples mentioned previously about complex systems can be also seen as self-organizing systems. Whether a system is considered as self-organizing does not depend only on the system, but also on the observer (Gershenson & Heylighen, 2003). There are several advantages in using the concept of self-organization in system design. While designing self-organizing systems, the designers focus on the behavior of the components and their interactions, so that through their dynamics, together they perform the system function without directly designing it. Since components interact constantly, it can be said that they are constantly searching for solutions. When a problem changes, the system adapts to the new situation, modifying its functionality.

An example of adaptation through self-organization has been proposed for traffic light coordination (Gershenson, 2005; Cools et al., 2007; Gershenson & Rosenblueth, In Press). Instead of trying to blindly predict when an average flow of vehicles should arrive at intersections, each traffic light gives preference to the streets with higher demand. In this way, vehicles on streets with lower demands will wait a bit longer, increasing the probability that more vehicles will join those that are waiting. This leads to the formation of vehicle "platoons". Once platoons reach a certain size, they will be able to trigger a green light before reaching an intersection, thus preventing platoons from stopping, unless other platoons or pedestrians are crossing at that moment. This platoon formation also leaves free spaces between platoons, allowing other platoons to flow with little interference. With simple local rules and without a direct communication between traffic lights, an adaptive synchronization is promoted, which adjusts itself to the immediate traffic conditions. The system adapts at the same *scale* at which the problem changes. The self-organizing method brings considerable improvements, reducing waiting times by more than 50%, saving not only time but also money, fuel, and pollution.

Another example of the benefits of self-organization can be seen in a recent proposal to regulate public transportation systems (Gershenson, 2011b). An algorithm that responds locally to passenger demand and vehicular intervals is able to perform even better than the theoretical optimum.

## *Language*

One of the main obstacles to adopt a novel scientific paradigm is our language. The way in which we speak, write, and describe things determines how we understand them. Newtonian dogmas find their roots in Platonic and Aristotelian language.

In the Greco-Latin worldview, which has dominated "western" cultures, *one* absolute truth is assumed. From this perspective, the mission of science is to "discover" the truths of the world. This presupposition becomes evident in classical logic, which includes the principle of the excluded middle (something is true or false, but not something else) and the principle of non-contradiction (something cannot be true and false at the same time). Classical logic, as well as traditional science, has been very useful, especially in closed systems.

Nevertheless, the truth of any proposition depends on its *context* (Gershenson, 2002b). This fact can be generalized from Gödel's (1931) incompleteness theorem. Gödel proved that in any formal system, such as mathematics, there are statements that cannot be proven. The root of this "problem" lies in the fact that axioms of a formal system cannot be proven from within that system, precisely because axioms are presupposed. This is relevant, because if axioms change, statements can change their truth value. For example, the statement "parallel lines never intersect" is true within Euclidean geometry. In fact, this is one of its axioms. However, there are other geometries, which do not consider this axiom, in which the statement is false, since parallel lines do intersect at the infinite. This can be visualized projecting the plane on a (Riemann's) sphere: if two parallel lines are projected on a sphere, these intersect on the opposite side of the sphere. This condition of formal systems leads to the "silly theorem problem": for any silly theorem (e.g. 1+1=10), there are infinite sets of axioms for which the silly theorem is true (e.g. use base 2). However, in practice this problem is trivial, because experience tells us which axioms are *useful*. Nonetheless, it should be noted that there is no set of "true" axioms. There are axiom sets over which formal theories can be based. Depending on the uses of the formal theory, others can be chosen. For example, Boolean algebra can be based on a single axiom (Wolfram, 2002). Still, proving theorems based on a single axiom can be more complicated that with other axiomatic systems.

Another example can be seen with Newton's laws, which were considered absolute truths, rulers of the universe. Still, at very small or very large scales, they do not apply. It is not that they are "wrong". Newton's laws apply to a certain context, and their common usage demonstrates their efficacy.

Within language, people have attempted to expel ambiguity by formalization, e. g. Tarski (1944). However, language is by nature ambiguous. It is better to understand contradictions (Priest & Tanaka, 1996; Gershenson, 1999) than ignoring them.

The limits of our descriptions can be illustrated with the following example. Assume there is a sphere, half white and half black. If the sphere can be seen only from one perspective, actually a circle will be perceived. Of which color is the circle? The answer will depend on the perspective from which the sphere is observed. The circle might be white, black, half and half, 10% white and 90% black, etc. (See Figure 3).

Of which color *is* the circle? The "right" answer will change depending on the perspective of the observer. Averaging answers will not be closer to the "truth", since it is highly probable that there are more observers from certain perspectives and less from others. In this case, we know beforehand that we are describing a sphere, not a circle, and that the sphere can be rotated and examined from a multitude of perspectives. However, all phenomena that we describe can have more and more features and dimensions.

One can never state that a phenomenon has been described completely, since our perceptions and descriptions are limited and finite. But one can always encounter in natural phenomena novel properties, interactions, relations, and dimensions. Just like with the sphere, we cannot describe completely any phenomenon, since our descriptions are limited and phenomena are not.

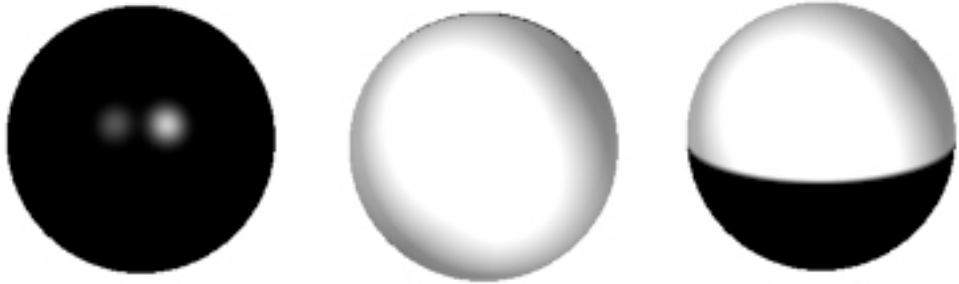

Figure 3. What *is* the color of the circle? It depends on the perspective from which the sphere is observed.

Does this imply that all hope of understanding the world should be abandoned? Certainly not. But we should be aware that our descriptions, if they are "correct", they would be so only within a determined context. There is no risk of "wild subjectivism", since our contexts are socially constructed. In other words, we reach agreements. What should be accepted is that there are no absolute truths, that the world changes, that our descriptions of the world also change, and that these changes have a limited predictability. We should take advantage of this dynamism instead of ignoring it or hopelessly trying to get rid of it.

Another example can be seen with colors. The color of an object can change depending on the illumination under which it is observed. In darkness, all objects are black. Behind rosy lenses, all objects are of a rose hue. Again, there is no risk of "radical relativism", because even when there might be more than one description for the same phenomenon, we can agree on the context under which the phenomenon is described and decide over its properties under a *shared context*. This leads us to reflect over the difference between the model and the modeled.

## *The model and the modeled*

Plato's myth of the cavern illustrates the presuppositions and aspirations of classical science, which are embedded in our language as well. Plato describes a cavern, where people are chained and can only see a wall. Behind them, different objects are found, which project their shadow on the wall. People can only see shadows. Plato writes that these people fool themselves, since they do not perceive reality. Philosophy (and later science) is the method to discover *the* truth. Philosophers can break their chains and see the reality in full color outside the cave, not only shadows.

Before going further, a distinction between the model and the modeled should be made. Models are *descriptions* of modeled phenomena. As such, models depend on the observer. Since there are no observations independent of observers, nor descriptions independent of a describer, there cannot be a

"direct" access to phenomena. Just by tagging them with a name, we are simplifying them to a description, a generalization, a model. This implies that even "breaking the chains", what people will see outside Plato's cave will not be *the* reality. It will be a *different description* of reality. It cannot be proven that this or any other description is the "correct" one, since the usefulness of descriptions depends on the purpose for which they are used. In other words, the only we can perceive is "shadows".

Humanity has always aspired for perfection. In science, this translates into seeking absolute truths. In engineering, this translates into faultless systems. We should admit our limits, our lack of perfection, and that our engineered systems are also limited and imperfect. These limits are natural and inherent, not a defect, since infinite potentialities cannot be contained within our finitude.

## *Conclusions*

We have illustrated how adaptation is essential to solve problems with non-stationary spaces. However, this does not imply that prediction should be neglected, as it is useful and desirable. Nevertheless, prediction should be complemented with adaptation and used with caution, considering its limits carefully. One of the accomplishments of the scientific study of complex systems is to show that the perfect control of open systems is utopic. If there are interactions, there will be a certain degree of unpredictability. This demands modesty and consideration when building systems and solving complex problems. The implication is that there will always be novel problems. The best we can do is to be prepared and expect the unexpected. It is not only desirable to have robust systems, so that they do not "break" because of perturbations. We should also give our systems a certain degree of creativity to be prepared to face the unknown.

## *Acknowledgements*

I am grateful to Xavier Martorell, Àngels Massip, Guillermo Santamaría, and the Fundación "la Caixa" for their invitation and organizational efforts. Héctor Zenil helped me to clarify certain doubts.